\input{aipcheck} 
 
\documentclass[ 
    ,final            % use final for the camera ready runs 
  ] 
  {aipproc} 
 
\layoutstyle{6x9} 
 
\begin{document} 
 
\title{Modeling of the spectral energy distribution 
 of the cataclysmic variable TT Ari 
 and evaluation of the system parameters} 
 
 
\author{K. V. Belyakov}{ 
  address={Kazan Federal University, Kremlevskaya str. 18, 42008 Kazan, Russia} 
} 
 
\author{V. F. Suleimanov}{ 
  address={Insitute for Astronomy and Astrophysics, Kepler Center for Astro and Particle Physics, Eberhard Karls University, Sand 1, 72076 T\"ubingen, Germany}, 
altaddress={Kazan Federal University, Kremlevskaya str. 18, 42008 Kazan, Russia}, 
email={suleimanov@astro.uni-tuebingen.de} 
} 
 
\author{E. A. Nikolaeva}{ 
  address={Kazan Federal University, Kremlevskaya str. 18, 42008 Kazan, Russia} 
} 
\author{N. V. Borisov}{ 
  address={Special Astrophysical Observatory, 369167 N. Arkhyz, Russia} 
}

\begin{abstract} 
The spectral energy distribution (SED) of the TT~Ari system, which is well 
 known from published IUE and optical photometric observations, was 
 modeled by a steady-state accretion $\alpha$-disc around a white 
 dwarf. Parameters of the system were derived from 
 time-resolved optical spectral observations in the bright state that we obtained in Sep.~1998.  The 
 radial velocity semi-amplitude of the white dwarf (33.8 $\pm$ 2.5 km 
 s$^{-1}$) and corresponding mass function ($f(M)$ = 5.5 $\pm 1.2~ \times 
 10^{-4}~ M_{\odot}$) were derived from the motion of the emission 
 components of Balmer lines. The mass ratio $q$ ($\approx$ 0.315) was 
 evaluated from the fractional period excess of the superhump period 
 over the orbital period $\varepsilon$ ($\approx$ 0.085), and a 
 secondary mass range ($0.18 - 0.38 M_{\odot}$) was estimated from the 
 orbital period. Therefore, the white dwarf mass range is 
 $0.57 - 1.2 M_{\odot}$ and the inclination angle of the system to the line 
 of sight is 17 -- 22.5 degrees. The adopted distance to the system is 
 335 $\pm$ 50 pc. 
 To fit the observed SED it is necessary to add a thermal spectrum with 
 $T \approx 11600$ K and luminosity $\approx 0.4 L_{\rm d}$ to the 
 accretion disc spectrum. This combined spectrum successfully describes 
 the observed Balmer lines absorption components. Formally the best fit 
 of the HeI 4471 line gives minimum masses of the components  ($M_{\rm 
 RD}$ = 0.18 $M_{\odot}$ and $M_{\rm WD}$ = 0.57 $M_{\odot}$), with the 
 corresponding inclination angle $i = 22.^{\circ}1$ and mass-accretion 
 rate  $\dot M = 2.6 \times 10^{17}$ g s$^{-1}$. 
\end{abstract} 
 
\maketitle 
 
%%%%%%%%%%%%%%%%%%%%%%%%%%%%%%%%%%%%%%%%%%%% 
%% MAINMATTER 
%%%%%%%%%%%%%%%%%%%%%%%%%%%%%%%%%%%%%%%%%%%% 
 
\section{Introduction} 
 
TT Ari is a bright (V$\approx$10.$^m$8) anti-dwarf nova with orbital period $P_{\rm orb}$=0.$^d$13755  
\citep{tors:85}. The photometric period varies from $P_{\rm ph}$=0.$^d$1329 \citep{smak:69}  
to $P_{\rm ph}$=0.$^d$14926 due to accretion disc precession \citep{skil:98}. In the bright  
state, optical spectra of the system show broad Balmer absorption lines with narrow central  
emission peaks together with HeII $\lambda$4686 and CIII/NIII $\lambda$4645 emissions  
\citep{sta:01,wu:02}. \cite{gans:99} determined the spectral class of the secondary  
(M3.5$\pm$0.5) and obtained a distance to the system $d$ = 335$\pm$50 pc. 
 
Here we present the spectroscopic observations of TT Ari and the estimation of the system  
parameters using spectral energy distribution (SED) modeling.

\section{Observations} 
 
\begin{figure} 
  \includegraphics[height=.25\textheight]{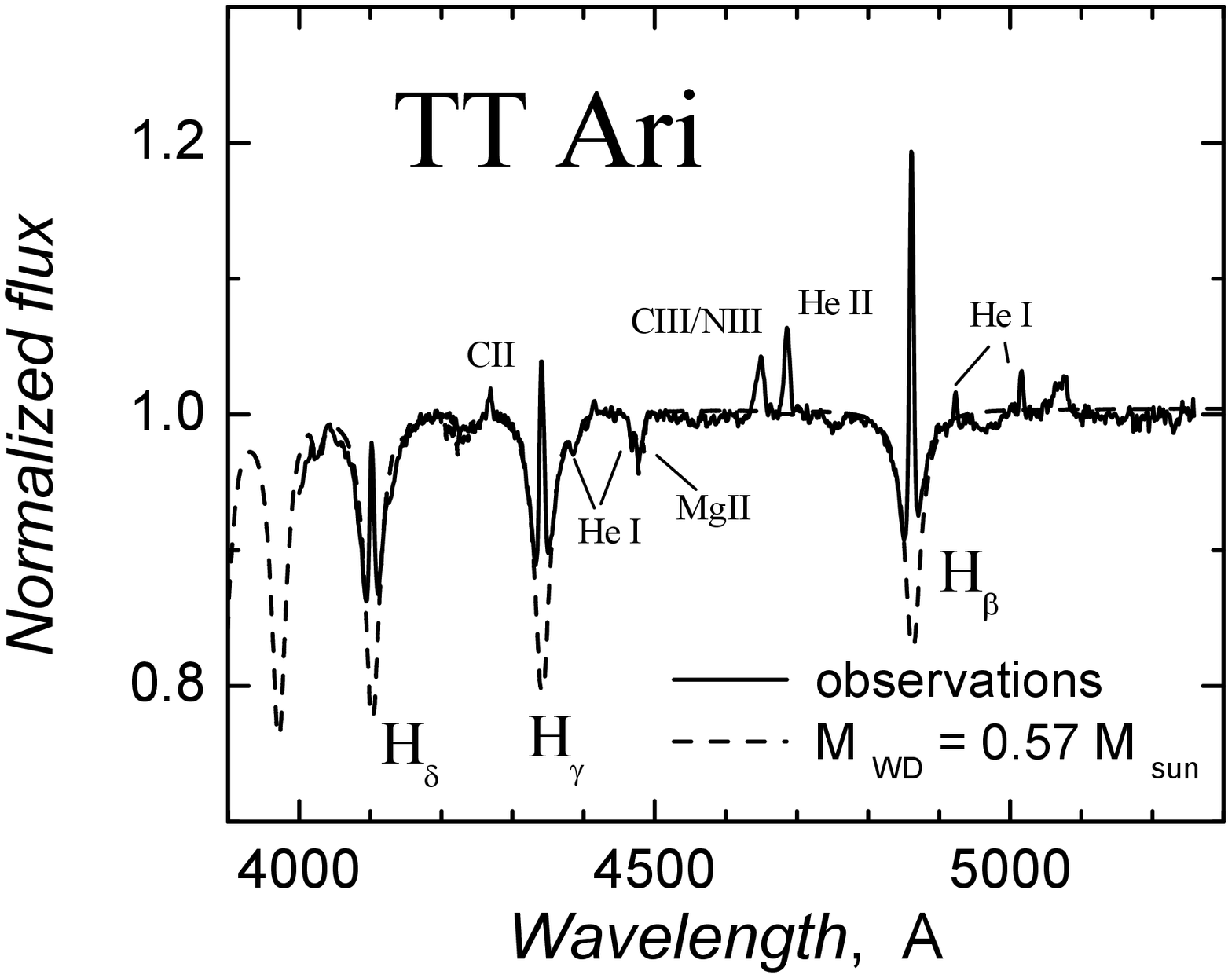} 
  \includegraphics[height=.25\textheight]{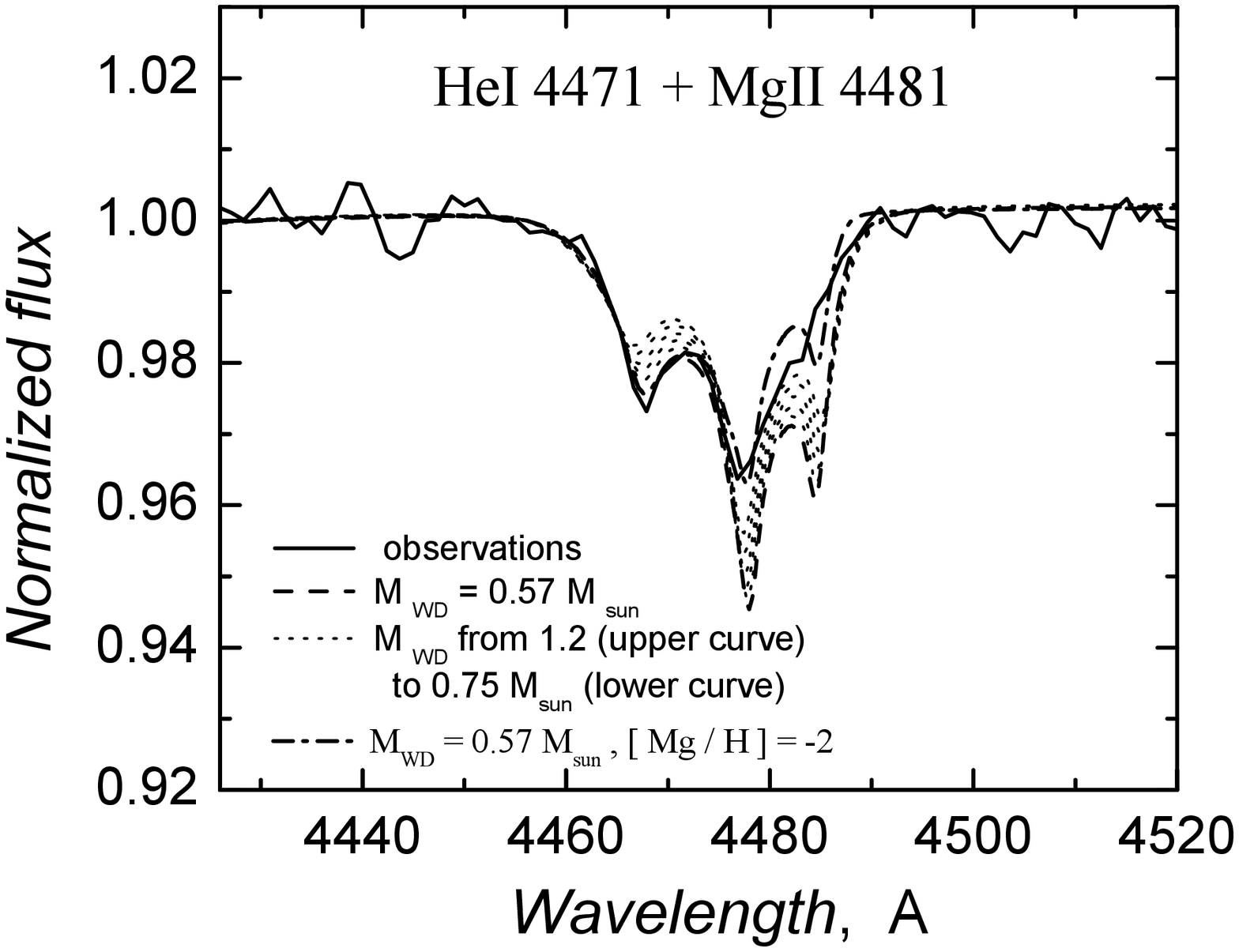} 
  \caption{ \label{v2sfig2} 
Comparison of the model spectrum with the additional black-body component 
 (dashed curves) with the observed spectrum (thick solid curves). } 
\end{figure}

Spectroscopic observations of TT Ari were carried out on Sept. 13-14, 1998, by the 6--meter 
 telescop BTA of the Special Astrophysical Observatory with 
the long-slit spectrometer SP-124, which gives a  $\Delta\lambda$ = 2.9 
\AA~ resolution in the wavelength region 4000--5300 
\AA. 180 consecutive spectra with the same exposure time of 60 s were obtained.  
The average normalized speectrum is shown in Fig.\,\ref{v2sfig2}. 
 
\begin{figure} 
  \includegraphics[height=.25\textheight]{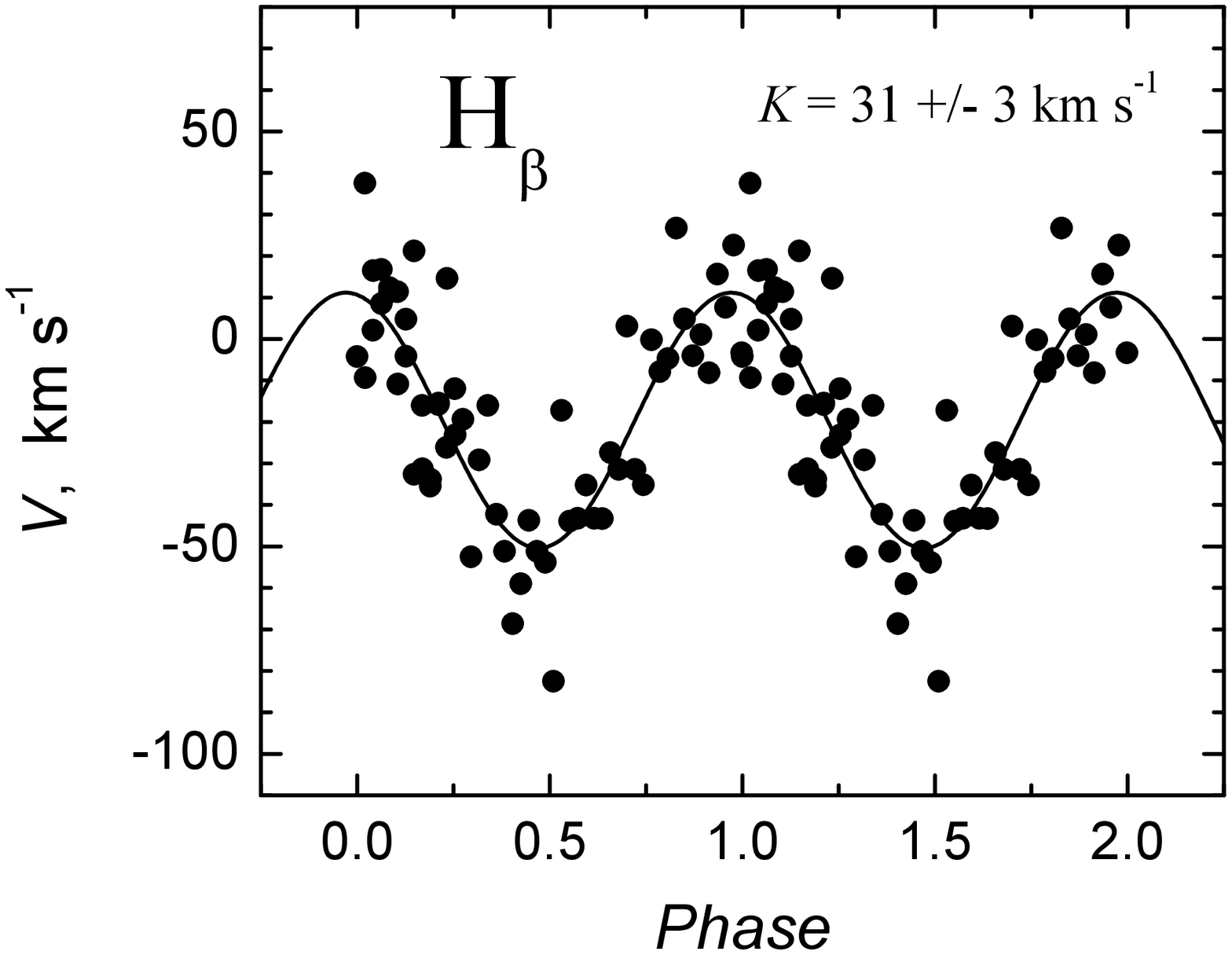} 
  \includegraphics[height=.25\textheight]{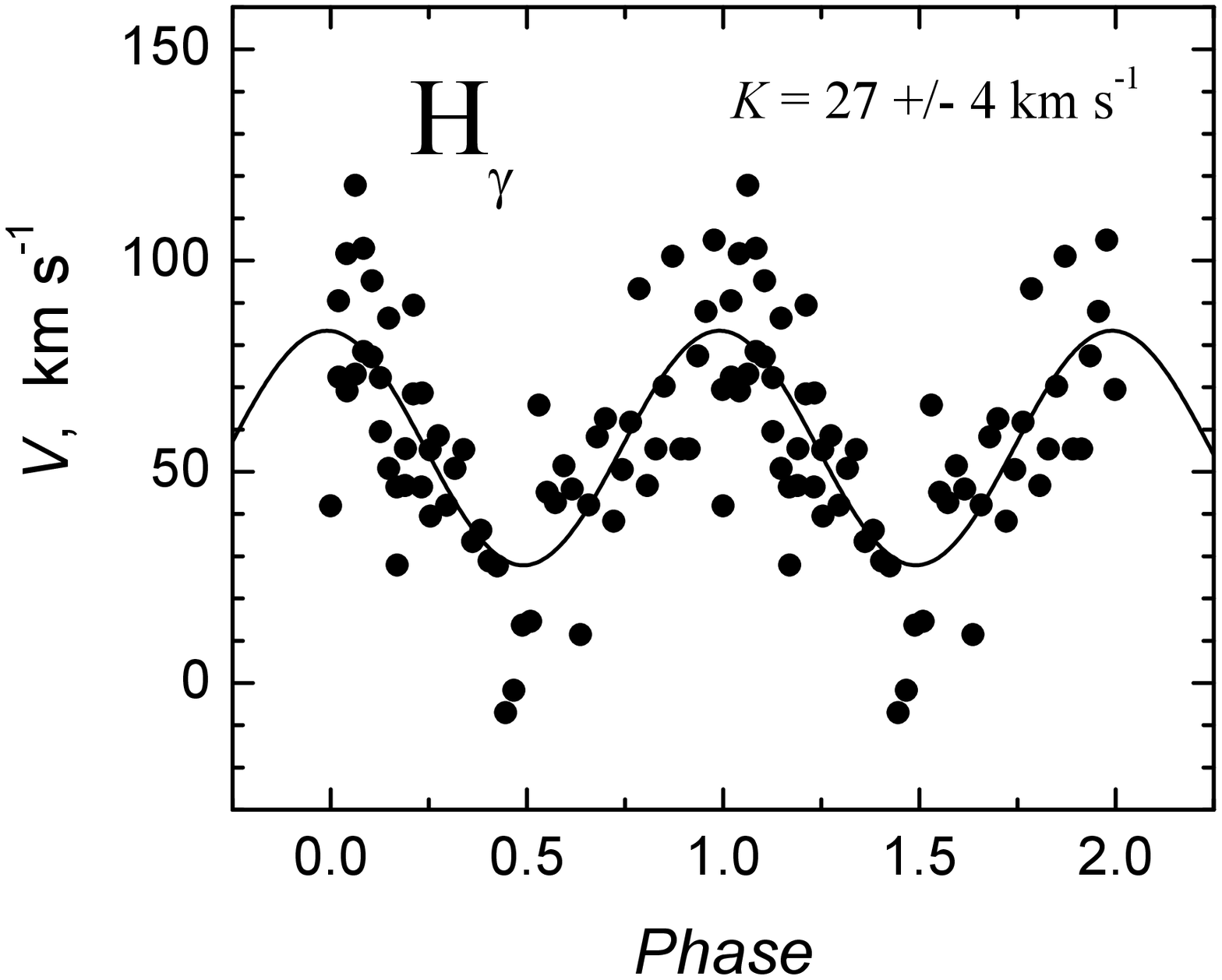} 
  \caption{ \label{v2sfig1} 
The radial velocity curves. The sines, which fit the 
radial velocity curves, are shown by solid curves. 
} 
\end{figure}

The radial velocities of the white dwarf were measured using the narrow emission cores of  
hydrogen lines. Dependences of the radial velocity on the orbital phase for H$_{\beta}$ and  
H$_{\gamma}$ are shown in Fig.~\ref{v2sfig1}.  
 
\section {Estimation of system parameters} 
 
We adopted $K$ = 33.75 $\pm$ 2.5 km s$^{-1}$ from our radial velocity measurements  
and other authors$^,$ results \citep{sta:01,wu:02}. Using this value, we derive the  
corresponding red dwarf mass function $f(M) = 5.5\pm 1.2~ \times 10^{-4}~ M_{\odot}$.   
 
The mass of the secondary star can be evaluated from the orbital period using the  
mass-radius relation for main sequence stars: $M_{\rm RD} = 0.18 - 0.38 M_{\odot}$  
\citep{rap:01}, $M_{\rm RD} \approx 0.26 M_{\odot}$ \citep{knig:06}, and  
$M_{\rm RD} \approx 0.24 M_{\odot}$ \citep{patt:05} .  
 
The mass ratio $q=M_{\rm RD}/M_{\rm WD}$ can be found from the fractional period  
excess $\epsilon = (P_{\rm ph}-P_{\rm orb})/P_{\rm orb}$ 
using the $\epsilon - q$ relation \citep{patt:05} 
\begin{equation} 
  \epsilon = 0.18\,q +  0.29\,q^2.  
\end{equation} 
This relation gives $q \approx$ 0.315, because $\epsilon \approx$ 0.085 for TT Ari.

\section{SED modeling} 
 
We tried to employ the SED of TT Ari to obtain an additional limitation on the  
TT Ari parameters. To model of the accretion disc spectra our code was used  
\citep{sul:92,sul:96}. In this code, stellar atmosphere spectra with solar  
chemical composition and corresponding $T_{\rm eff}$ and $\log g$ are taken  
as local spectra of the disc. The external disc irradiation is not taken into  
account. The accretion disc model spectra were computed for six models with  
parameters presented in Table\,\ref{v2st2}. The mass accretion rates were obtained  
using the observed flux at 1460 \AA~ (IUE,\cite{verb:87}), and the adopted distance $d=335$ pc. 
 
\begin{figure} 
  \includegraphics[height=.26\textheight]{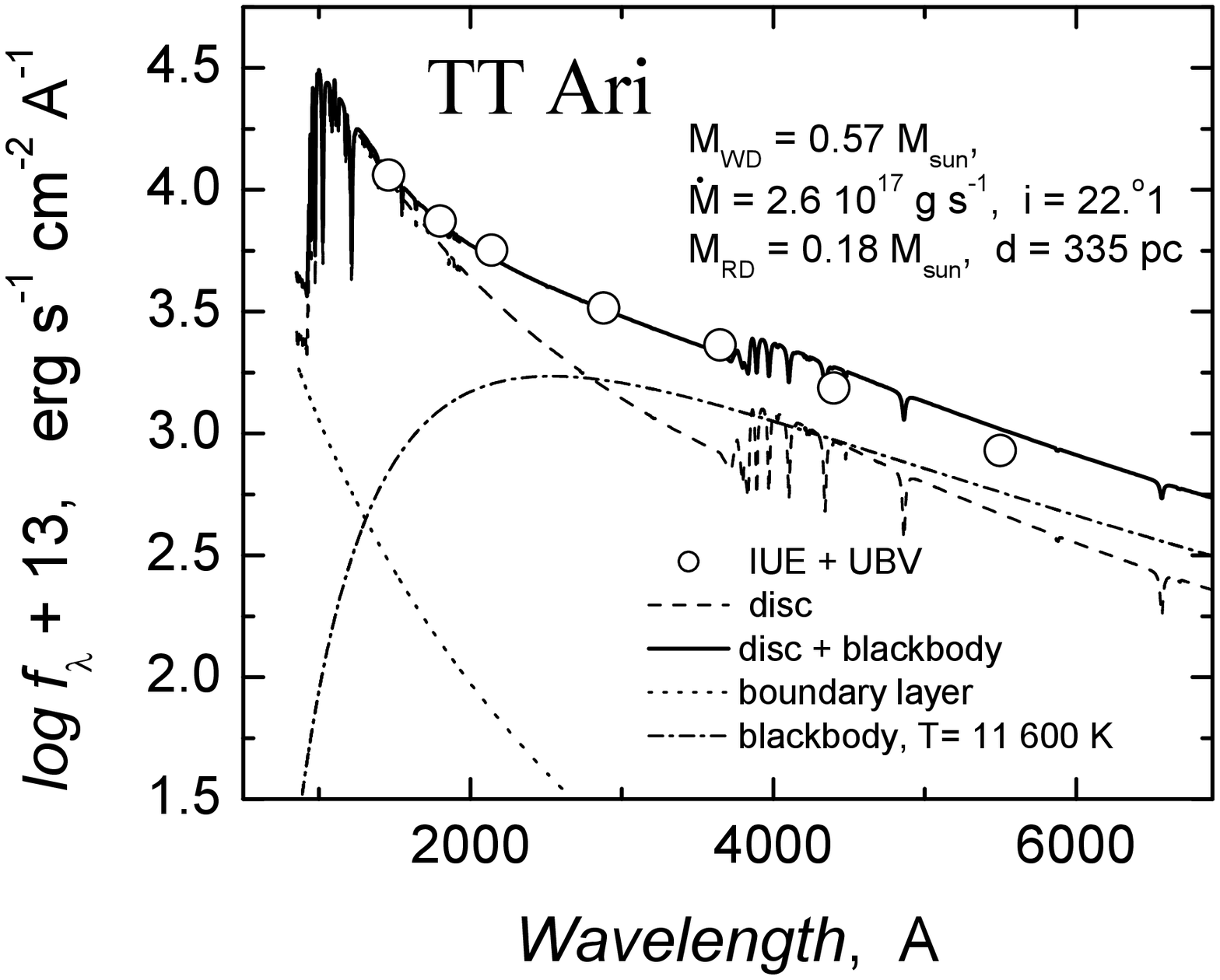} 
  \includegraphics[height=.26\textheight]{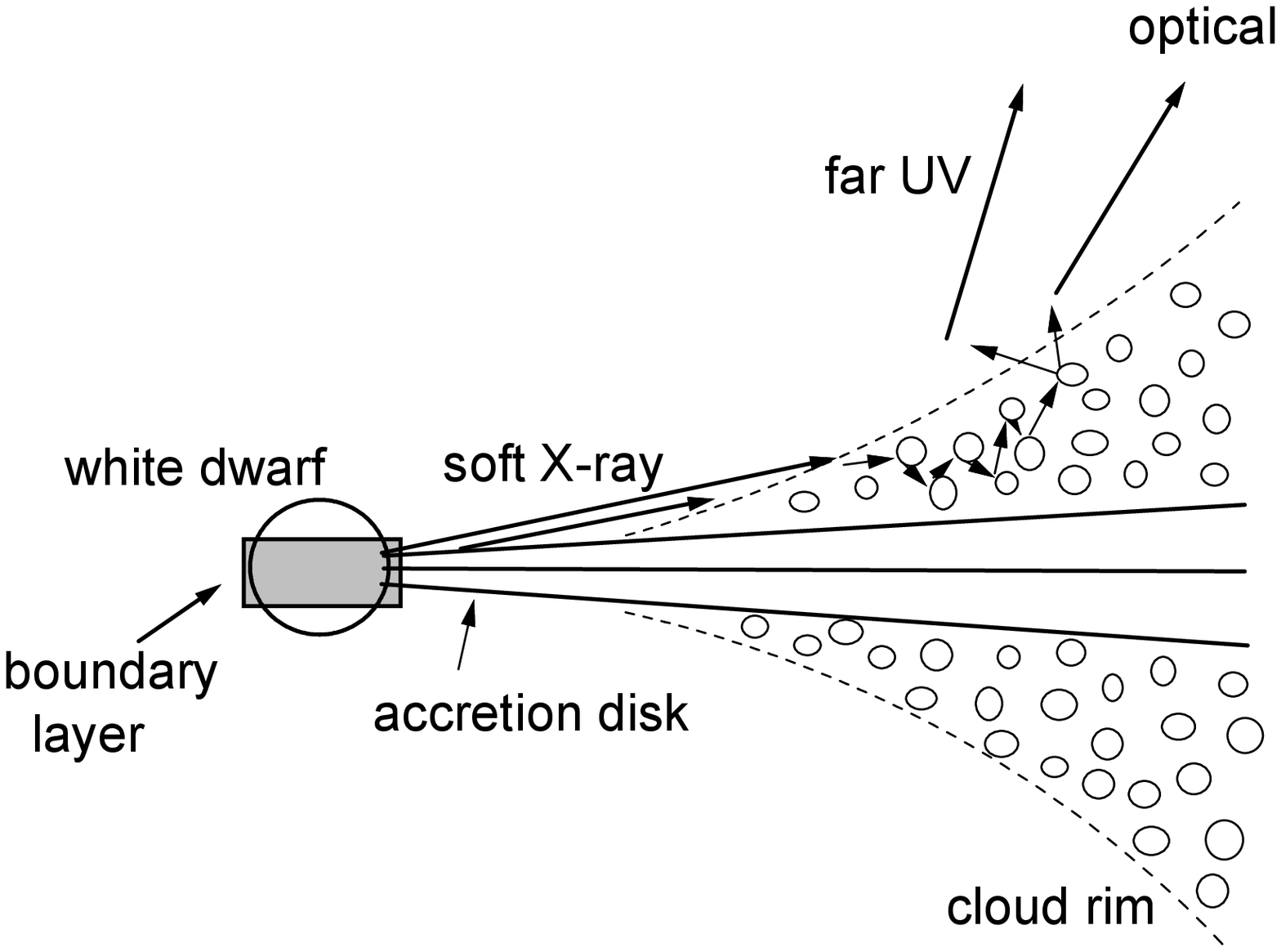} 
  \caption{ \label{v2sfig3}  
{\it Left:} Results of the spectral energy distribution modeling. {\it Right:}  
The qualitative picture of the cloud rim above an accretion disc. 
} 
\end{figure}

\begin{table} 
%\begin{center} 
\begin{tabular}{c|cccccc} 
\\ 
$M_{\rm RD}/M_{\odot}$ & 0.18 & 0.22 & 0.26 & 0.30 & 0.34 & 0.38 \\ 
\hline 
$M_{\rm WD}/M_{\odot}$ & 0.57 & 0.75 & 0.83 & 0.95 & 1.08 & 1.20 \\ 
\hline 
$i$, deg & 22.1 & 20.6 & 19.4 & 18.5 & 17.7 & 17.0 \\ 
\hline 
$\dot M/10^{17}$ g s$^{-1}$ & 2.59 & 1.91 & 1.48 & 1.18 & 0.94 & 0.76 \\ 
\end{tabular} 
\caption{Parameters of TT Ari accretion disc models.} \label{v2st2} 
%\end{center} 
\end{table} 
The accretion disc spectrum alone cannot describe the observed SED  
\citep{verb:87,be:94}, see Fig.\,\ref{v2sfig3}\,(left panel). This problem was first  
mentioned by \cite{wade:88}. Formally, the relatively good agreement  
with the observed SED can be obtained adding a blackbody spectrum with  
$T$ = 11\,600 K and luminosity $L_{\rm bb} \approx$ 0.4 $L_{\rm disc}$.  
This additional flux can arise due to irradiation of the outer disc by  
the central disc part and a boundary layer. But a huge outer disc  
half-thickness $H/R \approx$ 0.4 is necessary for this (at a radiation  
reprocessing efficiency $\eta$ = 0.3). There is a similar problem for  
the optical flux of supersoft X-ray sources \citep{dist:96} and can be  
also explained by a system of optically thick clouds above the outer  
disc \citep{sul:03}, see Fig.\,\ref{v2sfig3}\,(right panel). 
 
The model spectra with the blackbody component satisfactorily describe  
the widths of absorption wings of Balmer and HeI lines (Fig.\,\ref{v2sfig2}).  
The best description of the HeI 4471 line gives the pair  
$M_{\rm RD}$ = 0.18 $M_{\odot}$ and $M_{\rm WD}$ = 0.57 $M_{\odot}$.  
The red wing of this line is blended by the MgII 4481 line. A very  
small ($\approx$ 1\% of solar) magnesium abundance is suggested  
to explain the shallow depth of this line.  
 
\section{Conclusions} 
 We conclude that the observed TT Ari SED and the absorption components  
of  Balmer and HeI lines can be described by the accretion disc radiation  
with an additional blackbody ($T$ = 11\,600 K, $L_{\rm bb}$ = 0.4 $L_{\rm disk}$).  
The system parameters range between $M_{\rm RD} = 0.18 - 0.38 M_{\odot}$,  
$M_{\rm WD} = 0.57 - 1.2 M_{\odot}$ and the inclination angle $i = 22.^{\circ}1 - 17^{\circ}$. 
 
\begin{theacknowledgments} 
   This work is supported by the 
Russian Foundation for Basic Research (grant  09-02-97013-p-povolzhe-a), and the DFG SFB\,/\,Transregio 7 ``Gravitational Wave Astronomy'' (V.S.). 
\end{theacknowledgments} 

\bibliography{v2suleimanov}
\bibliographystyle{aipproc}
 
\end{document}